\documentclass[twocolumn,backref, breaklinks, colorlinks]{aastex63} 
\hypersetup{linkcolor=blue,citecolor=blue,filecolor=cyan,urlcolor=magenta}
\usepackage{tabularx}
\usepackage{url}
\usepackage{graphicx}
\usepackage[T1]{fontenc}
\usepackage{ae,aecompl}
\usepackage{booktabs}
\usepackage{natbib}
\usepackage{longtable}

\usepackage{blindtext}
\usepackage{longtable}
\usepackage{tabu}
\usepackage{lineno}

\newcommand\ba{{\it BatAnalysis}}

\def\aap{A\&A} \def\apjl{ApJ}  
 \def\apjs{ApJS}

 \submitjournal{ApJ}

\begin{document}

\title{ BatAnalysis - A Comprehensive Python Pipeline for Swift BAT Time-Tagged Event Data Analysis}

\author[0000-0002-4299-2517]{Tyler Parsotan}
\affiliation{Astrophysics Science Division, NASA Goddard Space Flight Center,Greenbelt, MD 20771, USA.}

\author[0000-0001-7128-0802]{David M. Palmer}
\affiliation{Los Alamos National Laboratory, Los Alamos, NM 87544, USA}
\affiliation{New Mexico Consortium, Los Alamos, NM, 87544, USA}

\author[0000-0003-0020-687X]{Samuele Ronchini}
\affiliation{Department of Astronomy and Astrophysics, 
The Pennsylvania State University, 525 Davey Lab, University Park, PA 16802, USA}

\author[0000-0001-5229-1995]{James Delaunay}
\affiliation{Department of Astronomy and Astrophysics, 
The Pennsylvania State University, 525 Davey Lab, University Park, PA 16802, USA}

\author[0000-0002-2810-8764]{Aaron Tohuvavohu}
\affiliation{Cahill Center for Astronomy and Astrophysics, California Institute of Technology, Pasadena, CA 91125, USA}

\author[0000-0003-2714-0487]{Sibasish Laha}

\affiliation{Center for Space Science and Technology, University of Maryland Baltimore County, 1000 Hilltop Circle, Baltimore, MD 21250, USA.}
\affiliation{Astrophysics Science Division, NASA Goddard Space Flight Center,Greenbelt, MD 20771, USA.}
\affiliation{Center for Research and Exploration in Space Science and Technology, NASA/GSFC, Greenbelt, Maryland 20771, USA}

\author[0000-0002-7851-9756]{Amy Lien}
\affiliation{University of Tampa, Department of Chemistry, Biochemistry, and Physics, 401 W. Kennedy Blvd, Tampa, FL 33606, USA}

\author[0000-0003-1673-970X]{S. Bradley Cenko}
\affiliation{Astrophysics Science Division, NASA Goddard Space Flight Center,Greenbelt, MD 20771, USA.}
\affiliation{Joint Space-Science Institute, University of Maryland, College Park, MD 20742, USA}

\author[0000-0003-4348-6058]{Hans Krimm}
\affiliation{National Science Foundation 2415 Eisenhower Ave., Alexandria, VA 22314}

\author[0000-0001-9803-3879]{Craig Markwardt}
\affiliation{Astrophysics Science Division, NASA Goddard Space Flight Center,Greenbelt, MD 20771, USA.}

\correspondingauthor{Tyler Parsotan}
\email{tyler.parsotan@nasa.gov}


\begin{abstract}
	
The Swift Burst Alert Telescope (BAT) is a coded aperture gamma-ray instrument with a large field of view that was designed to detect and localize transient events. When a transient is detected, either on-board or externally, the BAT saves time-tagged event (TTE) data which provides the highest quality information of the locations of the photons on the detector plane and their energies. This data can be used to produce spectra, lightcurves, and sky images of a transient event. While these data products are produced by the Swift Data Center and can be produced by current software, they are often preset to certain time and energy intervals which has limited their use in the current time domain and multi-messenger environment. Here, we introduce a new capability for the \ba{} python package to download and process TTE data under an open-source pythonic framework that allows for easy interfacing with other python packages. The new capabilities of the \ba{} software allows for TTE data to be used by the community in a variety of advanced customized analyses of astrophysical sources which BAT may have TTE data for, such as Fast Radio Bursts (FRBs), Gamma-ray Bursts (GRBs), Low Mass X-ray Binaries (LMXB), Soft Gamma Repeaters, magnetars,  and many other sources. We highlight the usefulness of the \ba{} package in analyzing TTE data produced by an on-board GRB trigger, a FRB external trigger, a sub-threshold detection of the LMXB EXO 0748-676, and an external trigger of a GRB that BAT detected during a slew.
\\

\end{abstract}




\section{Introduction}\label{sec:intro}
The Neil Gehrels Swift Observatory \citep{gehrels2004swift} launched on November 20th, 2004 with three telescopes onoboard. These are the X-ray Telescope (XRT; \citep{burrows_XRT}), the Ultraviolet-Optical Telescope (UVOT; \citep{roming_UVOT}), and the Burst Alert Telescope (BAT; \cite{barthelmy_BAT}). The Swift BAT was designed to detect and localize Gamma-Ray Bursts (GRBs) and has significantly advanced our understanding of these transient events. Besides GRBs, BAT has also triggered on a number of known sources, such as soft gamma repeaters, as well as other transients, such as tidal disruption events.

The BAT uses the coded mask technique to produce small localization regions ($\sim 3$ arc-minutes) and accurate background estimations while maintaining a large field of view ($\sim60^\circ \times 120^\circ$). When a transient triggers BAT on-board, the spacecraft autonomously slews such that the narrow field telescopes can observe the region of the sky where the transient was localized to. Additionally, the associated BAT data is downlinked to the ground. This data is the time-tagged event (TTE) data which records all measured photons around the time of the triggering event, providing the location on the detector plane where each photon was detected and each photon's energy. Besides on board triggers, TTE data can be obtained using the GUANO system which commands the spacecraft to save TTE data surrounding the time of an external trigger \citep{tohuvavohu2_guano}. Furthermore, TTE data is saved for ``failed'' triggers or events that didn't meet the triggering threshold for BAT to autonomously trigger and slew to the localized position of the transient. 

The BAT TTE data has been used in the literature for many different sources and analyses. The TTE data for BAT triggered GRBs were analyzed to produce the first, second, and third Swift Gamma-ray Burst (GRB) catalogs \citep{bat_grb_first_catalog, bat_grb_second_catalog, bat_grb_third_catalog} with pre-processed GRB lightcurves and spectra for the community to use. BAT TTE data is additionally used alongside XRT and UVOT data to provide a comprehensive view of GRBs from their prompt to afterglow emission \citep{burst_analyzer}. In depth analyses of the TTE data for GRBs has shown the presence of a lag–luminosity correlation for GRBs \citep{ukwatta_lag_lumi, parsotan2022grbreview} which can be used to constrain the physics of GRB jets. BAT TTE data has additionally been used to measure quasi-periodic oscillations from the magnetar SGR 1806–20 during a giant outburst \citep{palmer_qpo} and, more recently, from GRB 211211A \citep{chirenti_qpo}. On longer timescales, BAT TTE data was used alongside NICER data to study the outburst spectra from the magnetar Swift J1555.2-5402 \citep{Enoto_bat_tte_magnetar} over a course of a month while many years of TTE data was accumulated to analyze the fluctuations of the Crab and show that there is an anti-correlation between its flux and spectral photon index \citep{augustine_crab_anticorrelation}. 
The expanded capability of obtaining BAT TTE data for external triggers using the GUANO system has also allowed that data to be used to localize transient events using the NITRATES low-latency search, advancing the science return of many complimentary observatories \citep{delaunay_nitrates}. In the search for electromagnetic emission from gravitational wave sources, BAT TTE data has played a critical role in placing constraints on the expected rate of coincident gravitational wave-electromagnetic radiation detections from  binary Black Hole mergers and Neutron Star-Black Hole mergers \citep{ronchini_constraining_GW230529, raman_lvk_followup}. 

BAT TTE data plays a critical role in the current time domain and multi-messenger landscape, however its use has been limited by the tools that have been made to analyze this data type. The method to analyze BAT TTE data in the context of GRBs had been made public through the HEASoft \texttt{batgrbproducts} pipeline script, however this script does not allow for full flexibility in the different timing and energy parameters that a user may want to customize. Furthermore, this pipeline cannot handle the analysis of TTE data that is collected due to ``failed'' triggers or by the GUANO ground system. To make the full set of TTE data more accessible and modernize the analysis of this data in the context of the greater python environment, we have modified the \ba{} python package \citep{batanalysis, parsotan2023batanalysis} to be able to handle the analysis of TTE data alongside BAT survey data.

The structure of this paper is as follows. Section  \ref{TTE_general} outlines the general steps taken to analyze BAT TTE data. Section \ref{ba_code} discusses the \ba{} python package and how it can be used to analyze TTE data. Section \ref{Sec:discussion} presents the results of our \ba{} code as compared to prior analyses and shows new analyses that are possible with the software. Finally, in Section \ref{Sec:conclusions}, we conclude with future improvements of the \ba{} code. 

\section{BAT TTE Data} \label{TTE_general}
Here, we describe the general methodology of how BAT TTE data is collected and processed.

\subsection{Obtaining TTE Data} \label{producing_TTE}
The most well known method for obtaining TTE data from BAT is through an on-board detection of a transient. As BAT is detecting photons, the flight software constructs count rate lightcurves on different timescales and energy ranges. If the software detects an increase in rates it constructs an image of the sky and compares the location of the potential source of interest to an on-board catalog of known sources\footnote{This on-board catalog can be added to as new sources are detected. This prevents known sources from constantly re-triggering BAT unless the SNR of their detection exceeds some predetermined threshold.}. If there is a new source in the image, then this triggers the observatory to slew to the calculated position of the new source in the constructed image. As BAT is surveying the sky it is already producing these images of the sky with different integration times and scaling them, which are referred to as scaled maps. An alternative way for BAT to trigger is to detect a new source within these scaled maps greater than some signal-to-noise ratio (SNR) threshold. Thus, BAT can trigger based on a rate trigger, where an increase in count rate compared to the background meets a SNR threshold, or the detection of a new source in a scaled map of the sky. In either case, TTE data starting from before the trigger time to after the trigger time will be saved and downlinked; usually this is 10-15 minutes worth of TTE data. Furthermore, it is important to highlight that BAT can only trigger on a transient when the spacecraft is not slewing.

A capability that is not as well known is BAT's ability to save TTE data around some ``failed'' or subthreshold trigger. This is an event that the onboard software detects as meeting the on-board rate trigger threshold but does not surpass the on-board imaging SNR threshold. As a result, no actual trigger will occur where the observatory will slew. However, in this situation the on-board software will save a small amount of TTE data around the time of the detected rate increase. This amount of data is correspondent to $\sim 10$ seconds. This situation can crop up if a transient is simply not bright enough in the image domain or if a source listed in the on-board catalog does not produce a transient event that exceeds it's preset SNR threshold to trigger BAT.   

Finally, \cite{tohuvavohu2_guano} developed the GUANO ground system, which permits BAT TTE data to be captured for external triggers such as those from the Fermi Gamma-ray Burst Monitor \citep{fermi_gbm}. The GUANO system ingests the time of the external trigger and sends a command to BAT telling it to save some portion of TTE data surrounding the trigger time before it gets overwritten on-board\footnote{The status of these commands and the GUANO system can be found at: https://www.swift.psu.edu/guano/. This webpage also lists the different types of external triggers that motivated the GUANO data dump}. The commanding is successful $>90\%$ of the time which means that either 200 or 90 seconds of TTE is later downlinked to the ground.

\subsection{Processing TTE Data} \label{processing_TTE}

\begin{figure*}
    \centering
    \includegraphics[width=\linewidth]{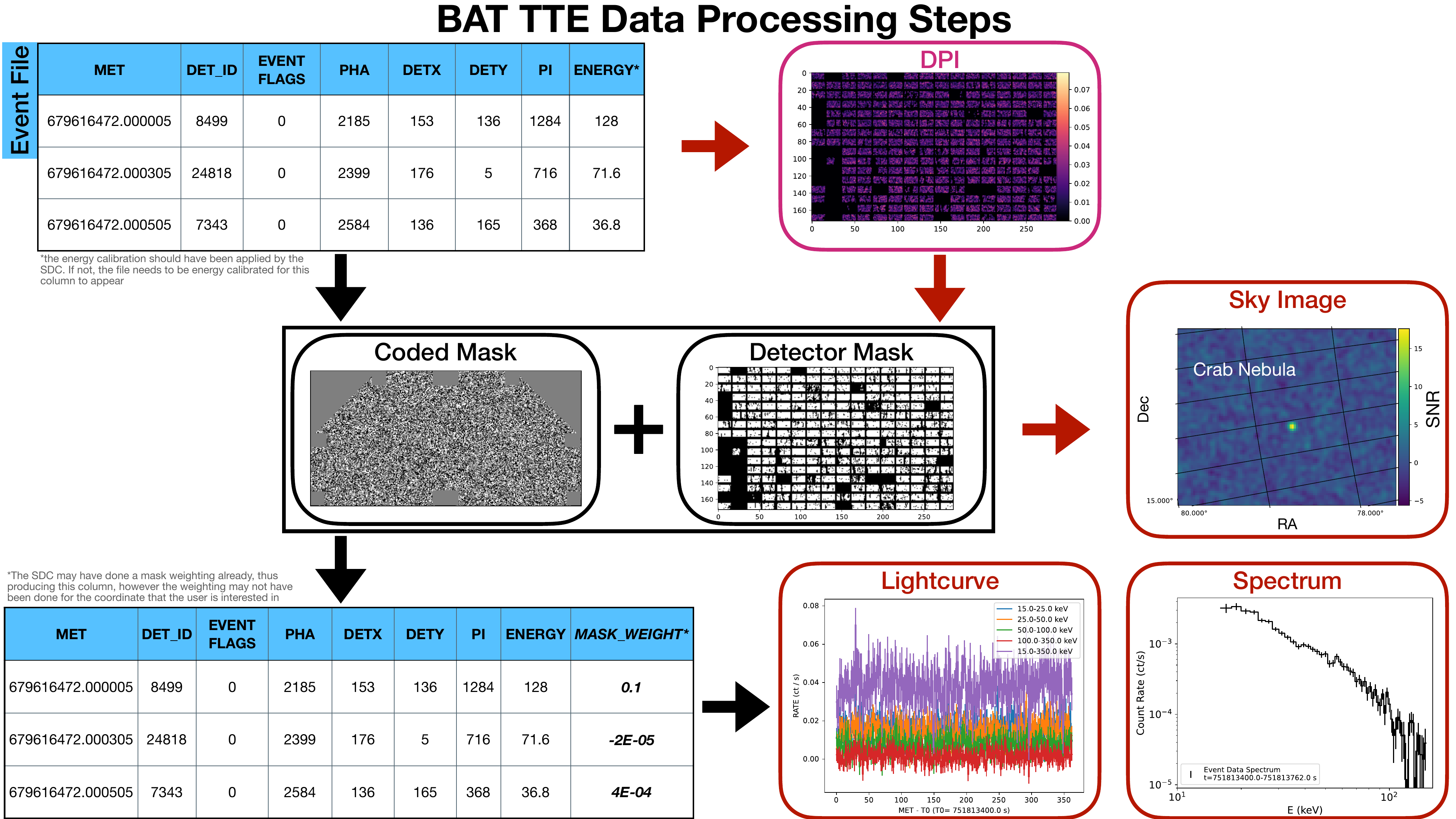}
    \caption{A flowchart outlining the different steps taken to obtain various high level data products from BAT TTE data corespondent to a Crab calibration observation. Starting with an event file that has been energy calibrated (which is typically done at the Swift Data Center), a user can create a DPI. The DPI is cross correlated with the coded mask to produce the sky image. This path is represented by the red arrows in the flowchart. To produce the mask-weighted lightcurve and spectra for the source of interest, the event file has to be modified by appending the file with a new column that contains the calculated mask weights. Then, the lightcurves and spectra can be calculated by binning the event data into their respective time and energy bins. These steps are represented by the black arrows on the flowchart. As a note, the Swift Data Center (SDC) typically does the energy calibration for the event data but if this is not done, the users will need to apply the energy calibration themselves. Similarly, the SDC may already calculate the mask weights for a position on the sky, which will be reflected in the event file. If the location on the sky that was used in calculating the mask weights is not the coordinates of the source that the user is interested in, then they will need to redo the mask weighting themselves. }
    \label{fig:tte_data_processing}
\end{figure*}

The processing of the BAT TTE data is implemented in the \texttt{batgrbproduct} HEASoft script. This script processes TTE data that was collected by an on-board trigger using predetermined energy and temporal ranges for spectra and lightcurves. Here, we briefly describe the steps followed by this script, which is also shown in a high level flow chart in Figure \ref{fig:tte_data_processing} \footnote{Additional information can also be found at: https://swift.gsfc.nasa.gov/analysis/bat\_swguide\_v6\_3.pdf and https://swift.gsfc.nasa.gov/analysis/BAT\_GSW\_Manual\_v2.pdf}.  While we mention of some aspects of coded-mask instrumentation and data analysis techniques, a more in-depth discussion of the various techniques can be found in \cite{braga_coded_review} and \cite{goldwurm_coded_review} and the references therein. 

TTE data is a list of events that have been detected on the BAT detector plane, where an event refers to something that generates a count on the detector plane such as a photon or a cosmic ray. This list contains the event's energy and the location on the detector plane where it was measured. The location is especially critical as it allows for the coded mask weighting to be calculated and images to be constructed.

The HEASoft script \texttt{bateconvert} ensures that the TTE data has the appropriate energy calibration applied to it. It uses BAT housekeeping data to apply the appropriate gain and offset calibrations. This task is typically done by default by the Swift Data Center (SDC) when it is processing the triggered TTE data from the BAT.

An additional complication is the fact that some portion of BAT's 32,768 detectors may be disabled or actively noisy. These detectors can be turned off by the instrument team to address operational issues or they can be turned off by the flight software if it identifies a given detector as being noisy. This data is reflected in the global quality mask, which is updated in the HEASARC's Calibration Database (CALDB), and the enable/disable map, which is automatically included in the observation ID dataset that is downloaded. Sometimes, detectors become noisy during a trigger. These detectors can be identified by creating a detector plane image (DPI) from the event data and using the HEASoft script \texttt{bathotpix} to identify which detectors are noisy during the trigger. All of these different maps of disabled or noisy detectors are combined to give us the detector quality mask which is incorporated into the analysis of the BAT TTE data. 

A TTE dataset from a triggered event will be processed by the SDC and, by default, the events will be weighted by the location of the event that triggered BAT. This location is the position identified by the onboard flight software which gets downlinked via the Tracking and Data Relay Satellite System (TDRSS). The mask-tagging algorithm identifies which rays of light from the location are able to propagate to the detector plane without being attenuated by the coded mask. Detectors that receive no photons from the position on the sky get a weight, $w=-1$, while detectors that are fully illuminated get assigned a weight, $w=1$. Partially illuminated detectors are assigned weights between $-1$ and $1$ based on the proportion of the detector that is illuminated. This weighting process transforms the Poisson statistics associated with counting photons on the detector plane to Gaussian statistics. Furthermore, it means that the resulting lightcurve and spectra produced from the weighted event data correspond to just the source of interest. In other words, the weighting effectively removes the background due to diffuse X-ray emission as well as other sources in the field of view\footnote{The removal of background sources with this method is not complete as photons from other sources in the FOV can still be detected at a given location on the detector plane where the associated weight is non-zero. Thus, there is technically still some contamination from other sources with the mask-weighting method.}. Changing the location on the sky for which mask weighting is performed will eventually produce a new background subtracted lightcurve and spectra for that new position in the sky. The HEASoft \texttt{batmaskwtevt} script does the mask weighting for a TTE dataset, taking a number of corrections into account including the number of active detectors and the geometrical corrections needed for a given source position with respect to the instrument boresight. \added{This is shown in Figure \ref{fig:tte_data_processing} by following the black arrows on the left of the figure connecting the energy calibrated event file to the mask weighted event file.} Not only does this script modify the event file, but it outputs the auxiliary ray tracing file which traces the position of the coordinate of interest on the detector plane as the spacecraft slews. This information is necessary for analyzing spectra, especially those that overlap with a slew.

Constructing the lightcurves and spectra for the single position in the sky is simply a sum of all the events multiped by their weights \footnote{There is an implicit unit of `per detector' that gets applied to the lightcurves and spectra units when the weights are applied such that the units of counts/s really mean counts/s/detector.}. The HEASoft \texttt{batbinevt} script does this operation for both lightcurves and spectra\added{, and is represented in Figure \ref{fig:tte_data_processing} by the black arrow that connects the modified event file to the mask-weighted lightcurve and spectra}. The spectra require additional information related to the location of the sky coordinate with respect to the detector boresight and the systematic error on the spectra. The geometric information of the sky coordinate position on the detector plane is taken from the auxiliary ray-tracing file that was previously created and inserted into the spectrum file using the \texttt{batupdatephakw} HEASoft script. The HEASoft script \texttt{batphasyserr} obtains the BAT systematic error for each energy channel from CALDB and appends it to the spectral file. Finally, to construct a detector response matrix (DRM)  we use the \texttt{batdrmgen} HEASoft script which should be used to construct a response for each spectrum where the source of interest is at a different location within the BAT field of view, for example each spectrum that is constructed during a slew. 

Sky images can also be constructed from the BAT TTE data. This is done by creating DPIs for the time interval of interest and using the \texttt{batfftimage} HEASoft script to deconvolve the DPI with the BAT coded aperture mask taking a number of coded mask and geometric corrections into account. The \texttt{batfftimage} script can also be used to construct the partial coding map of the sky, which represents the percentage of the BAT detector plane that is exposed to a given point on the sky. \added{These steps are represented as the path that follows the red arrows in Figure \ref{fig:tte_data_processing} connecting the event file, the DPI, the coded mask and detector mask, and the resulting sky image.}

The \texttt{batgrbproduct} script processes a given TTE dataset that was produced by an onboard trigger and produces lightcurves, spectra, detector responses, and images following the methodology outlined above. The default weighted, or background subtracted, lightcurves produced by \texttt{batgrbproduct} are binned into either a single (15-350 keV) energy channel or 4 energy channels (15-25, 25-50, 50-100, and 100-350 keV) in time bins of 64 ms or 1 s. Additional lightcurves that are generated are a 15-350 keV lightcurve in 4 ms timebins, an unweighted 15-350 keV lightcurve binned into 4 ms timebins, and a single and 4 energy channel Bayesian block lightcurve constructed using the \texttt{battblocks} HEASoft script. The output of \texttt{battblocks} also includes estimates of the trigger's time to accumulate 90\%, 50\% and 100\% of the photons from the triggering source ($T_{90}$, $T_{50}$, and $T_{100}$ respectively),  as well as the peak 1 s time interval, and the background time intervals. The spectra that \texttt{batgrbproduct} creates by default have 80 energy channels, between 15-150 keV\footnote{Technically, BAT has a non-coded response up to 500 keV but the 15-150 keV energy range is recommended for conducting spectral fits with BAT event data since the coded mask becomes transparent above $\sim 150$ keV. Since the spectra are constructed by mask-weighting the TTE data this technique only works when the mask is opaque to the detected radiation, in the 15-150 keV band. Additionally, the detector responses are not calibrated above 150 keV \citep{delaunay_nitrates} further complicating the spectral fitting above this energy.}, and they are constructed for the various time intervals identified with \texttt{battblocks}. Additional spectra are constructed for the pre-slew, post-slew, and slew time periods. The detector responses that are created correspond to the pre-slew and post-slew time periods. Users can only use these responses when fitting models to spectra that were constructed for these pre- and post-slew time periods. \texttt{batgrbproduct} also produces DPIs and sky images for the same pre- and post-slew time intervals in both single and 4 energy channels. DPIs and images are constructed for the time interval prior to any detectable emission from the source, which is denoted at the ``pre-burst'' time interval. Finally, partial coding maps are constructed for the pre- and post-slew time intervals. 

The \texttt{batgrbproduct} script comprehensively takes a number of calibrations and coded-mask corrections into account to produce the high-level lightcurves, spectra, detector responses, DPIs, and sky images which users can use directly. However, the preset time and energy binning make the output incompatible with joint spectral analysis studies or timing studies of the lightcurves. Furthermore, this pipeline does not handle processing subthreshold trigger TTE data or GUANO TTE data. 

\section{The \ba{} Code} \label{ba_code}

In this section we outline the \ba{} \citep{batanalysis} python package's TTE data analysis capability\footnote{This package is open source and is available on github at: https://github.com/parsotat/BatAnalysis \added{\citep{batanalysis}} }.
The \ba{} package allows a user to:
\begin{itemize}
	\item[1.] Download BAT triggered, subthreshold, and GUANO TTE data
	\item[2.] Process the BAT TTE dataset
	\item[3.] Construct custom lightcurves, spectra, and images 
	\item[4.] Search images for flaring sources 
	\item[5.] Conduct spectral fitting
\end{itemize}
The \ba{} package is a wrapper for HEASoftpy which allows for the identical processing of BAT TTE data as what is outlined in Section \ref{processing_TTE}. 
Thus, the files produced by the processing of the TTE data are also identical to what is produced by the HEASoft scripts. The \ba{} package allows for the high level BAT TTE data products to be integrated into the greater astronomical data analysis python environment. In this section we elaborate on each of the prior highlighted capabilities of the software. 

The documentation included in the \ba{} github repository goes into the details of how to utilize the software to accomplish the tasks outlined in this section. Furthermore, the codes that were used to produce the results in Section \ref{Sec:discussion} provide examples of how to use the python package to process the different TTE datasets.

\subsection{Downloading BAT TTE datasets} \label{sec:download}
  
Prior methods of querying and downloading BAT datasets required using the High Energy Astrophysics Science Archive Research Center (HEASARC) for observations correspondent to a trigger that a user is interested in and then downloading the datasets through their web interface. Downloading GUANO or subthreshold trigger is less straightforward as these TTE data types are archived using different schemes\footnote{This archiving scheme is documented in the ``Overview of the Swift Data Archive'' document available at https://swift.gsfc.nasa.gov/archive/} and can be difficult to query based on time or position criteria. The \ba{} python package interfaces with the swifttools \citep{swifttools} python package, to allow for querying and downloading observation IDs based on different criteria. Users are able to search and download ``normal'' triggered TTE data, GUANO downlinked TTE data, and subthreshold TTE data. In particular, the \ba{} software takes into account the need for temporally associated energy calibration that typically has to be applied to subthreshold TTE data.

\subsection{Processing BAT TTE observations}\label{Sec:obs}

The steps for processing TTE data are identical to the ones outlined in Section \ref{processing_TTE}. As a result, it calls many of the HEASoftpy scripts that are needed to process the data including performing energy calibration, determining bad detectors, applying mask weighting, and constructing DPIs, lightcurves, spectra, and images.  Since these same HEASoft scripts are called in the background, the output files are identical to those that are produced using the steps previously outlined. Users can additionally pass a dictionary that has key/value pairs that correspond to the HEASoft parameters and the values that the user would like to set for those HEASoft script parameters. Overall, much of the documentation that is relevant for the various HEASoft scripts is still relevant for the \ba{} software. 

There are major advantages with processing data using \ba{}. The pythonic interface allows users to dynamically rebin DPIs, lightcurves, and spectra in energy and time while offering plotting interfaces that utilize matplotlib for quickly viewing these different products and eventually producing publication ready figures. Furthermore, the images that are generated by \ba{} can easily be projected onto a healpix map making them more compatible with modern astrophysical analyses and mosaiced to enhance any detected sources in images.

\subsection{Spectral Fitting} \label{Sec:spectral_fitting}
Once a user has created a PHA file with \ba{}, the appropriate corrections are automatically applied and an associated DRM gets automatically created. The PHA can be fitted within the \ba{} software through the use of pyXspec, similar to the capabilities that are available for spectra obtained for survey data. With pyXspec, \ba{} will fit a model to the spectrum utilizing Gaussian statistics. The default model is a \texttt{cflux*powerlaw} model that fits the spectra from 15-150 keV and users can pass in their own spectral models based on what they may expect. Additionally, the user can use the PHA and DRM files created by \ba{} in their own spectral fitting tools but special care may be needed to ensure that the appropriate systematic errors are properly applied. 

Using the spectral fittings, users can determine if the source is significantly detected following the same method used for survey data \citep{parsotan2023batanalysis}. A flux upper limit can be determined by scaling the background variation by the significance level that the user is interested in and then fitting the resulting spectrum to obtain the flux needed to make such a measurement \citep{parsotan2023batanalysis, Laha_2022_FRB}. 

As is highlighted by \cite{bat_grb_third_catalog}, the time period that a user would like to construct and fit spectra  may overlap with a spacecraft slew. In this case, \cite{bat_grb_third_catalog} constructed an averaged DRM over the time period where Swift is slewing. This averaged DRM can be constructed by taking the weighted average of many DRMs that were constructed on smaller time intervals. The weighting can be equal to the fraction of photon counts in the smaller time intervals where the individual DRMs are constructed. Using the \ba{}, the construction of such a time averaged DRM is possible and its use in spectral fittings is convenient.  

\subsection{Mosaiced Images}
\replaced{Similar to the mosaicing of survey data to increase the SNR of dim sources it is possible to mosaic the sky images constructed from BAT TTE data.}{It is possible to interpolate the images produced using BAT TTE data onto the sky and add them together, identical to what is done with BAT survey data. This technique can be utilized in a similar method to the dithering technique used in narrow field instruments, where images of a source which is moving through the FOV are stacked leading to an increased SNR of the source of interest.} In the case of GUANO downlinked TTE data, this capability also allows for searches of transient sources while BAT is slewing which is particularly important as Swift spends a larger portion of time slewing from source to source as a target of opportunity focused facility. When Swift is slewing the triggering is turned off for BAT which has an adverse effect on the number of on-board triggered GRBs since Swift has become more driven by target of opportunity observations. 

Mosaicing BAT TTE images is straightforward with the \ba{} package. Users first construct sky images as usual and then these sky images can be added together. Behind the scenes, the \ba{} is linearly interpolating each sky image onto a healpix map of a given resolution and then adding each healpix map together. The different sky images are added identical to the formalism used for survey data except there is no energy dependent off-axis correction which takes the width of the coded mask into account \citep{parsotan2023batanalysis}. The total flat exposure
for each pixel in the mosaic healpix map, $i$, is calculated as: 
\begin{equation}
E_{i}=\sum_k e_{i,k}
\end{equation}
where $E_{i}$ is the mosaic image's flat exposure in each pixel and $e_{i,k}$ is the flat exposure map of the $k^{\mathrm{th}}$ sky image that will be included in the mosaic once it has been interpolated on the healpix map. 
To calculate the partial coding vignetting mask we multiply the flat exposure by the partial coding once each sky image has been interpolated onto the healpix map and add the values in each pixel. This is given as
\begin{equation}
P_{i}=\sum_k e_{i,k}p_{i,k}
\end{equation}
where $P_{i}$ is the mosaic partial coding exposure map and $p_{i,k}$ is the partial coding map for the $k^{\mathrm{th}}$ image that is being mosaiced once it has been interpolated onto the healpix grid.

The standard deviation of the mosaiced image for energy band $l$, $\sigma_{i,l}$, is calculated by adding the inverse variances and then converting back to normal standard deviation. This is given as:
\begin{equation}
\sigma_{i,l}=\frac{1}{\sqrt{\sum_k {{v_{i,k,l}^{-2}}}}}
\end{equation}
where $v_{i,k,l}$ is the energy dependent standard deviation map of each individual image once it has been interpolated onto the healpix grid. 

To calculate the mosaiced flux sky image for energy band $l$, $S_{i,l}$, we add the flux sky images weighted by their associated inverse variance map. This is given as:
\begin{equation}
S_{i,l}=\sigma_{i,l} {\sum_k s_{i,k,l}\ v_{i,k,l}^{-2} } \label{sky_mosaic} 
\end{equation}
where $s_{i,k,l}$ is the flux sky image in each energy band that the user wishes to mosaic together interpolated on the healpix grids before being multiplied by $v_{i,k,l}$. 

These SNR in each pixel of the mosaic healpix map at a given energy can then be calculated simply as
\begin{equation}
    \textrm{SNR}_{i,l}=\frac{S_{i,l}}{\sigma_{i,l}}
\end{equation}

Sources in the mosaiced images can be searched based on the SNR of each pixel where typically SNR values $>7$ can be trusted. As an additional check for new sources, the healpix pixel's distance from known sources can be checked and compared to the full width half maximum size of the BAT's point spread function. This check ensures that potential high SNR pixels in the mosaic are distinct from known sources that have been previously detected by BAT. \cite{slew_survey_copete} outlines many of the details needed to be considered in mosaicing together images from TTE data with the goal of detecting and localizing transient sources.  

\subsection{Parallelized Analysis}
There are additional capabilities offered in \ba{} to allow users to expedite their analyses through parallelized functions. 
These functions allow users to perform the following operations in parallel: produce spectra, fit spectra, produce images, and mosaic images. While these functions are simplified to permit straightforward analyses in a parallelized manner, users can use them as templates for their own personal analyses with differing analysis requirements.  


\section{Applications of BAT TTE data}\label{Sec:discussion}
In this section, we validate the software by analyzing the on-board triggered GRB 050724, a short GRB with extended emission. Additionally, we show the capabilities of the \ba code as applied to BAT GUANO data coincident with a FRB external trigger from FRB 180916,  subthreshold trigger data that was collected for the LMXB EXO 0748-676, and BAT GUANO data that was saved due to the Fermi trigger of GRB 210706A\footnote{All the code for the examples shown here can be found in the notebooks subdirectory of the github repository: https://github.com/parsotat/BatAnalysis/tree/main/notebooks.}.

\begin{figure}
    \centering
    \includegraphics[width=\linewidth]{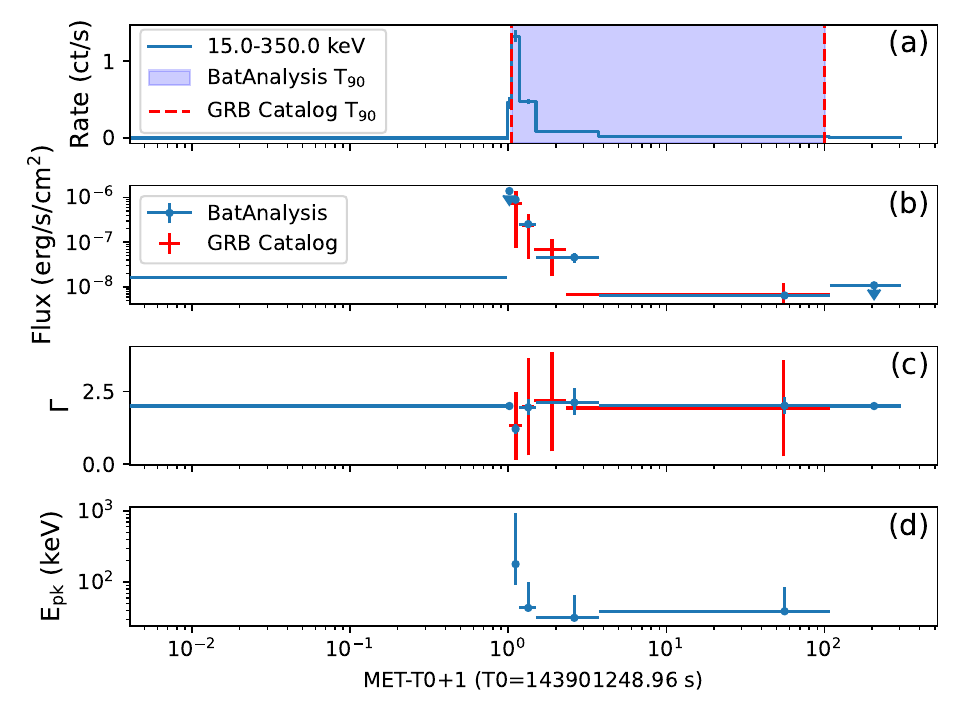}
    \caption{The mask-weighted lightcurve (with implied rate units of counts/s/detector) and fitted time-resolved spectral parameters for GRB 050724, including estimates of the spectral E$_\mathrm{pk}$. In panel (a), we plot the \ba{} Bayesian binned lightcurve and we highlight the measured  $T_{90}$ in the blue highlighted region. We also show the $T_{90}$ published in the GRB BAT Catalog for this event. In Panel (b) and (c) we show the \ba{} fitted spectral flux and power law photon index respectively in blue. We plot the spectral fits for the spectra provided in the GRB Catalog in red. In panel (d) we plot the estimated spectral E$_\mathrm{pk}$ \citep{sakamoto2009epeak}. To better highlight both the initial short pulse and the extended emission, the times are shifted by 1 second with respect to the trigger time. }
    \label{fig:grb_lc}
\end{figure}

\subsection{GRB 050724 - On-board triggered TTE Data}

The first science case that we use to exemplify the capabilities of \ba{} and verify the software is with the analysis of GRB 050724. This unique GRB is a short GRB with extended emission due to the short pulse that lasts for less than 2 seconds however the measured $T_{90}$ is significantly longer at $\sim 100$ seconds. BAT was triggered by this event on 2005-07-24 at 12:34:09.361840 UTC and promptly slewed to the coordinates determined on-board for this burst.

In Figure \ref{fig:grb_lc} we show the mask weighted lightcurve and the time resolved spectral parameters as a function of time. Both the lightcurve and the spectra are offset from the trigger time by 1 second in order to better show both the short pulse and the extended emission. In Figure  \ref{fig:grb_lc}(a), we show the Bayesian binned count rate lightcurve in the 15-350 keV energy range. We highlight the $T_{90}$ obtained by \ba{} with the blue highlighted region as well as the $T_{90}$ published for this GRB on the BAT GRB Catalog website \citep{bat_grb_third_catalog}. In Figure \ref{fig:grb_lc}(b) and Figure \ref{fig:grb_lc}(c) we plot the flux and the power law index obtained from fitting the time resolved spectra. The blue points show the quantities obtained by fitting the spectra using their associated DRMs produced by \ba{} with a \texttt{cflux*powerlaw} model for each timebin of the Bayesian binned lightcurve. The red points are the time resolved spectra obtained from the BAT GRB Catalog for this GRB. The time binning is slightly different and the spectra that are produced are fitted with a  \texttt{cflux*powerlaw} using the time averaged DRM that is available on the catalog webpage for this GRB. In Figure \ref{fig:grb_lc}(d), we use the \ba{} \texttt{cflux*powerlaw} spectral fit parameters to estimate the spectral peak energy, E$_\mathrm{pk}$, using the method outlined by \cite{sakamoto2009epeak}. We find that the \ba{} software is able to recover the duration of this GRB and the time resolved spectral parameters, albeit with smaller errors in the fitted parameters and a $\sim 1$ second difference in the start and end time of the timebins that encompass the extended emission. The increased precision of the \ba{} spectral parameters is due to the fact that \ba{} uses the DRM that is constructed for the same time interval that is used to construct the PHA file. This is an improvement over the spectral analysis presented in the GRB catalog where the time averaged DRM is used in all spectral fittings. This is only necessary for the last timebin of the extended emission phase where Swift is slewing. 

\begin{figure}
    \centering
    \includegraphics[width=\linewidth]{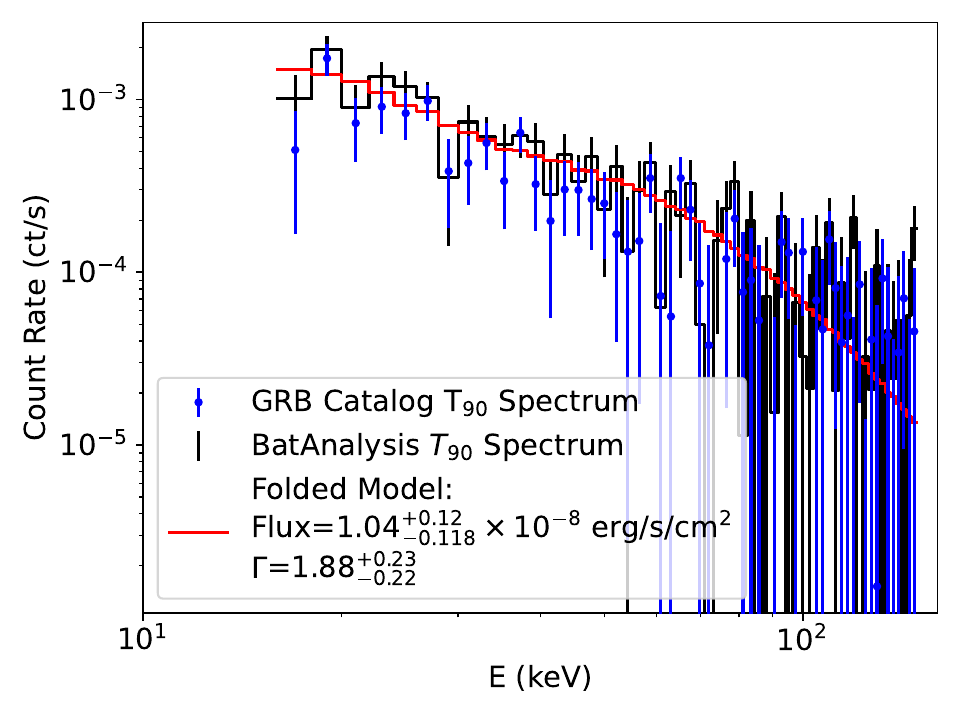}
    \caption{The \ba{} produced $T_{90}$ spectrum, plotted in black, agrees with the $T_{90}$ spectrum obtained in the BAT GRB catalog for this GRB, plotted in blue. Since this spectrum overlaps a time period where Swift is slewing we construct a time averaged DRM which we use in our spectral fitting. The best fit \texttt{cflux*powerlaw} model to the \ba{} spectrum is shown in red with the spectral parameters provided in the legend.}
    \label{fig:grb_spect}
\end{figure}

We also directly verify the spectra that are obtained within the $T_{90}$ time range in Figure \ref{fig:grb_spect}, which encompasses a time period when Swift is slewing. Here, we plot the \ba{} calculated 80 channel spectrum plotted for the 15-150 keV energy range  in black with the $1\sigma$ errors in each channel. We construct a time averaged DRM and use it to fit the \ba{} $T_{90}$ spectrum using a  \texttt{cflux*powerlaw} model and show the folded model spectrum in red. Finally, we plot the $T_{90}$ spectrum in the GRB catalog in the same 15-150 keV energy range in blue points. We can see that the \ba{} spectrum agrees well with the spectrum published in the BAT GRB catalog, verifying the correctness of the software.  

\subsection{FRB 180916 - GUANO TTE Data}\label{sec:guano}

The next science case that we use to showcase the capabilities of the \ba{} python package is analyzing GUANO downlinked TTE data correspondent with an external detection of FRB 180916. The \ba{} package makes it straightforward to process this data and eventually place flux upper limits on the FRB in the 15-150 keV energy range. 

\begin{figure}
    \centering
    \includegraphics[width=\linewidth]{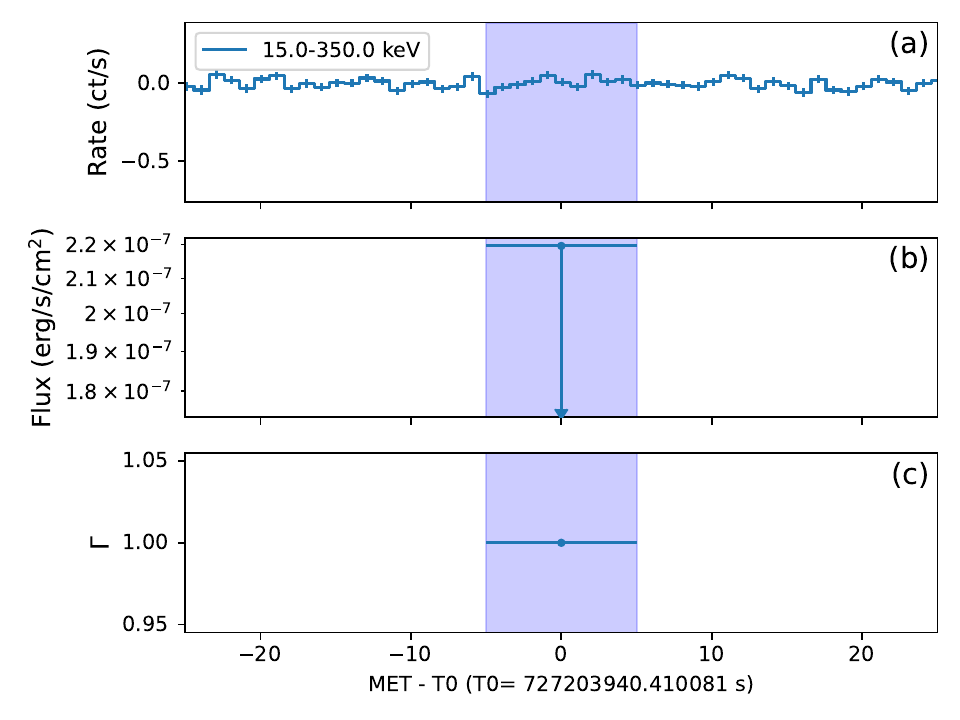}
    \caption{We plot the mask weighted lightcurve for the location of the FRB in the 15-350 keV energy range in panel (a) and also highlight, in light blue, the T0$\pm 5$ s time period where we place upper limits on the FRB emission. In panel (b) we show the flux upper limit obtained for the FRB in the time period of T0$\pm 5$ s. In panel (c) we plot the photon index, $\Gamma$, used to fit a \texttt{powerlaw} model to the upper limit spectrum to obtain the upper limit.}
    \label{fig:frb_lc}
\end{figure}

We select a GUANO dataset where the partial coding for the coordinates of the FRB is $\sim 19\%$. The dataset that we analyze is correspondent to the FRB trigger time on 2024-01-17 at 17:05:07.661808 UTC and the GUANO data is associated with Observation ID 01209406000. The results of the mask weighted lightcurve calculation and the flux upper limit that is obtained for the FRB is shown in Figure \ref{fig:frb_lc}. In Figure \ref{fig:frb_lc}(a), we show the mask weighted 15-350 keV lightcurve for the FRB, showing the 200 seconds of data that was saved by the GUANO system and the lack of a clear signal around the FRB trigger time. As as result, we use the \ba{} software to place upper limits on the FRB emission in the 15-150 keV energy range from T0-5 seconds to T0+5 seconds. We calculate the $5\sigma$ upper limit spectrum following the method outlined in section \ref{Sec:spectral_fitting} and fit a powerlaw with a photon index of 1. Our flux upper limit of $\sim 2.2 \times 10^{-7}$ erg/s/cm$^2$ is shown in Figure \ref{fig:frb_lc}(b) and the power law photon index used to obtain the upper limit is shown in Figure \ref{fig:frb_lc}(c).

\subsection{LMXB EXO 0748-676 - Subthreshold Trigger TTE Data}\label{sec:subthresh}
We use the subthreshold trigger that occurred with LMXB EXO 0748-676 to showcase the \ba{} package's ability to facilitate the finding and analysis of this type of TTE data. 

\begin{figure*}
    \centering
    \includegraphics[width=\linewidth]{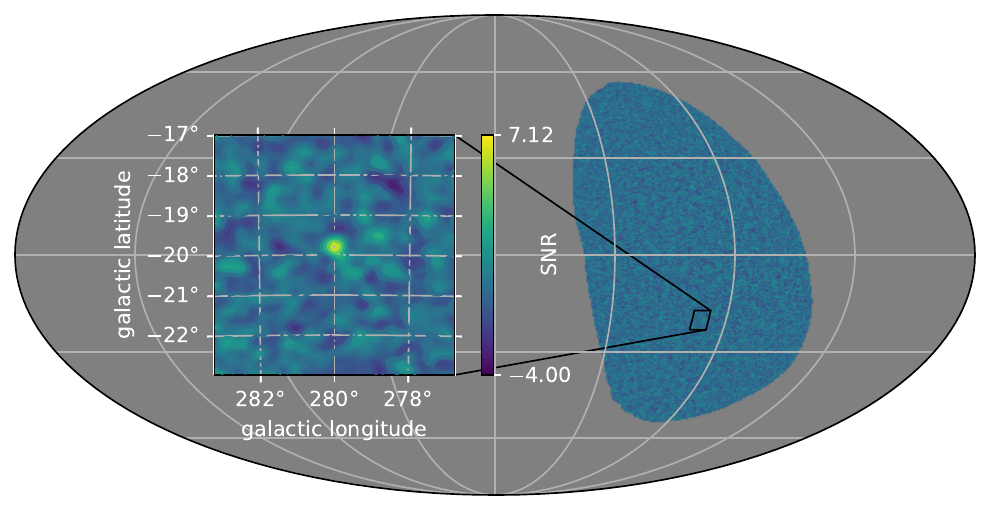}
    \caption{Using the subthreshold trigger TTE data, we show that there is a SNR excess at the location of EXO 0748-676. We plot the 15-25 keV SNR map of the sky, in galactic coordinates, as viewed by BAT at the time in a mollewide projection. We also plot a zoomed in view of the location of the LMXB, showing the SNR of the source which caused the on-board software to save the subthreshold TTE data.}
    \label{fig:exo_img}
\end{figure*}

In July 2024, there were a set of subthreshold triggers that were due to this source \citep{exo_atel}. Since this source is known and the outbursts did not meet the trigger thresholds for this source (except for one outburst which triggered BAT \citep{exo_trigger}) the subthreshold TTE data was saved by the onboard software. We analyze data for subthreshold trigger \#1245446, which occurred on 2024-07-29 at 12:05:16 UTC to show that this source is actually flaring, as is shown in Figure \ref{fig:exo_img}. Here, we show the BAT's field of view in galactic coordinates on a healpix map. The plotted quantity is the SNR map in the 15-25 keV energy band. The inset plot, shows the location of EXO 0748-676 and the SNR peak associated with the LMXB with SNR $\sim 7$.

\begin{figure}
    \centering
    \includegraphics[width=\linewidth]{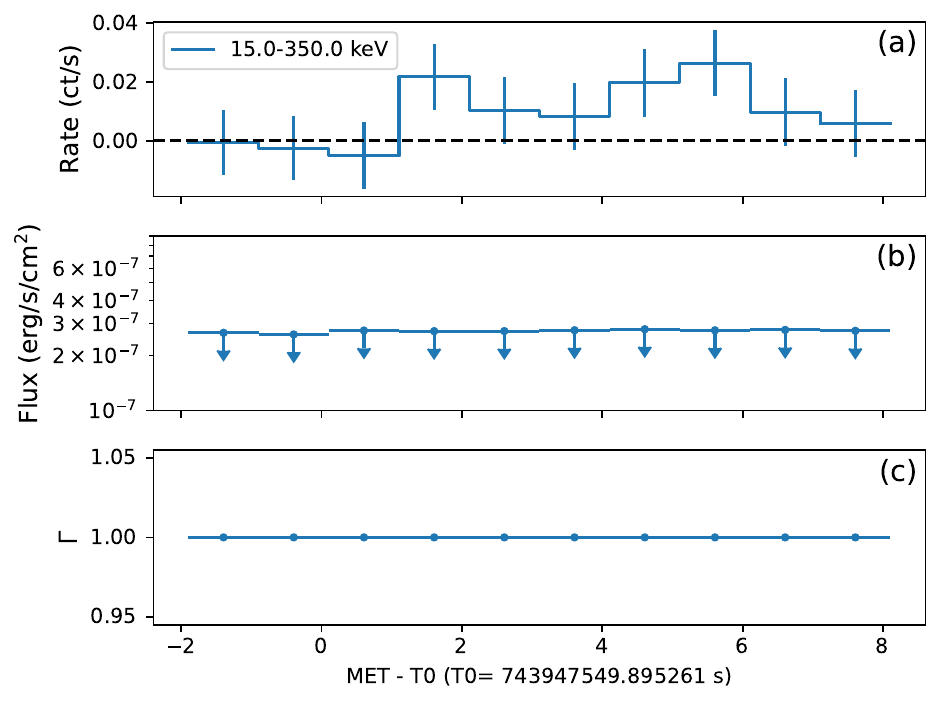}
    \caption{In panel (a) we plot the mask weighted lightcurve for EXO 0748-676 in the 15-350 keV energy range. In panel (b) we show the time resolved flux upper limits obtained for the LMXB in the time period of the subthreshold TTE data. In panel (c) we plot the photon index, $\Gamma$, used to fit \texttt{powerlaw} models to the upper limit spectra to obtain the upper limits.}
    \label{fig:exo_lc}
\end{figure}

In Figure \ref{fig:exo_lc}(a) we show the mask weighted lightcurve of the source. We can see a slow rise in the lightcurve for the data that is analyzed. We use the time resolved spectra to place flux upper limits on the LMXB in the 15-150 keV energy range, which is shown in Figure \ref{fig:exo_lc}(b). The photon index, $\Gamma$, used to fit the upper limit spectra with a \texttt{powerlaw} model is shown in Figure \ref{fig:exo_lc}(c). We find that the flux of the LMXB is $\lesssim 3\times 10^{-7}$ erg/s/cm$^2$ in  all 1 second time bins. 

\subsection{GRB 210706A - Slew Imaging Analysis}\label{sec:slew}

\begin{figure}
    \centering
    \includegraphics[width=\linewidth]{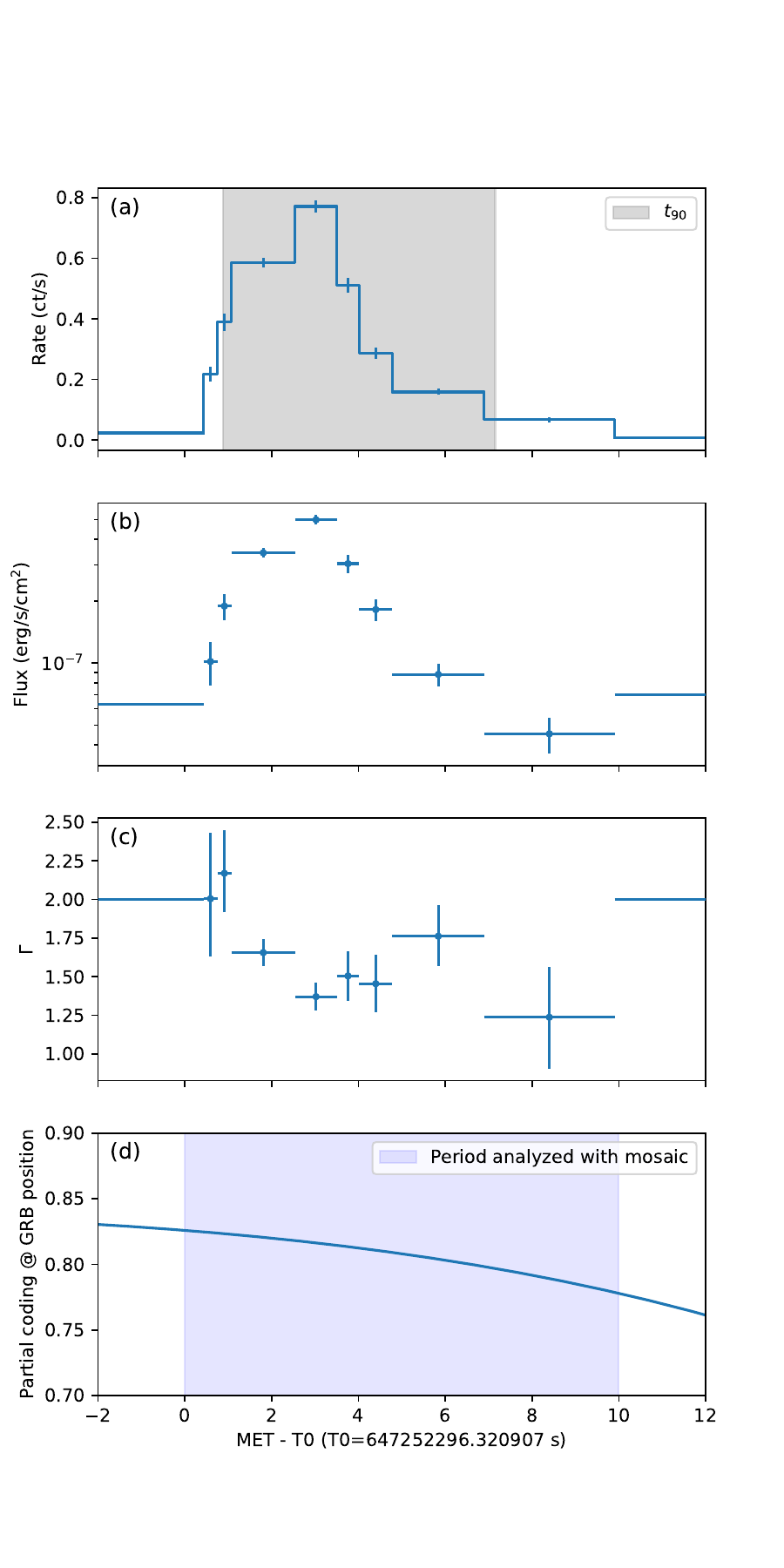}
    \caption{Results from the analysis of GRB 210706A which occurred while Swift was slewing. Panel (a): lightcurve rebinned using Bayesian blocks. The shaded gray area identifies the starting and ending time of the T$_{90}$ interval. Panel (b) and (c): flux and photon index evolution, from time-resolved spectral analyses using a \texttt{cflux*powerlaw} spectrum. Panel (d): temporal evolution of the partial coding evaluated at the position of the GRB showing the slew that Swift was performing at the time.}
    \label{fig:slew_lc}
\end{figure}

\begin{figure}
    \centering
    \includegraphics[width=\linewidth]{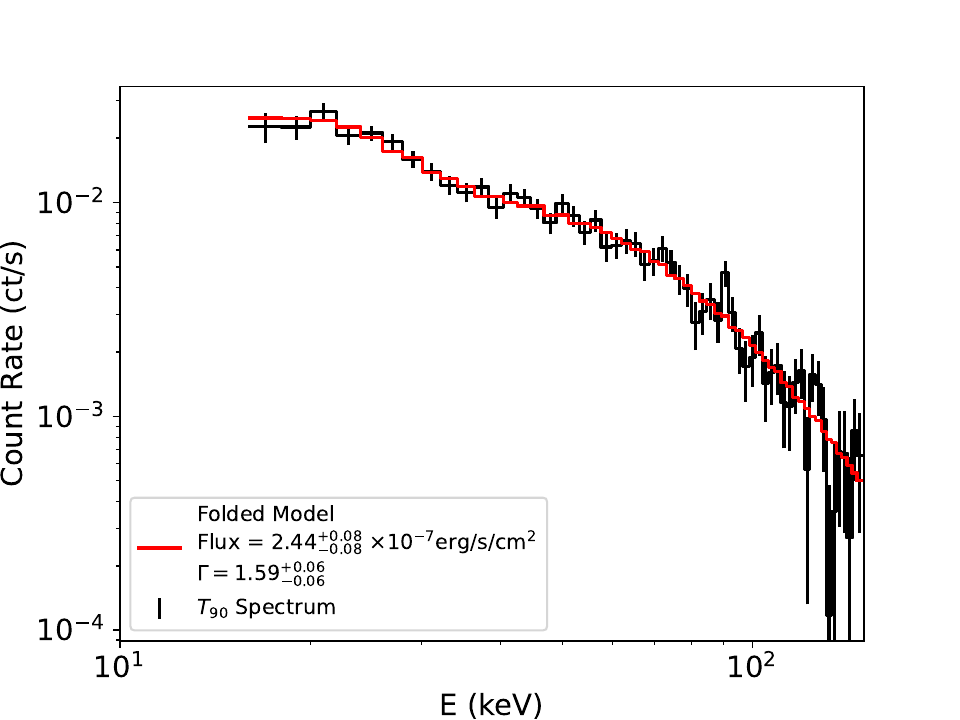}
    \caption{The time-integrated T$_{90}$ spectrum and the fitted \texttt{cflux*powerlaw} model of GRB 210706A. The spectral fit was conducted with a time averaged DRM.}
    \label{fig:spe}
\end{figure}

In this section we show that the \ba{} can be used to analyze a GRB that was detected in GUANO TTE data while Swift was slewing. We consider GRB 210706A, detected by Fermi on 2021-07-06 at 08:17:49.9 UTC. GUANO successfully downlinked 200 s of TTE data around the GRB trigger time. From an inspection of the attitude file, we can verify that the spacecraft was slewing at T0. To localize the GRB with the TTE data, we create a mosaiced image in the energy range of 15-350 keV for the time interval spanning $\Delta t$ = [T0,T0+10s], where the exposure of each image that gets mosaiced is 0.2 s following the method outlined by \cite{slew_survey_copete}. We find a source with SNR = 39.6 at RA = 311.96 deg, Dec = 13.29 deg, compatible with the position reported in the GUANO GCN \citep{slew} and in the XRT refined analysis GCN \citep{slew_ref}. After applying the mask weighting using these coordinates, we can extract the lightcurve and spectrum for this GRB. The background-subtracted, Bayesian binned lightcurve and temporal evolution of the partial coding at the GRB position are shown in Figure \ref{fig:slew_lc}. Using the same Bayesian block timebins, we computed the spectral evolution, using again a power law spectrum for each time bin, and the results are shown in panel (b) and (c) of Figure \ref{fig:slew_lc}. Here, the DRM constructed for each time interval was used in the fitting of each time resolved spectrum. Using the Bayesian blocks algorithm, we obtain a T$_{90}$ = 8.5 s, measured in the time interval $t_{GRB}$ = [-0.42, 8.09] s. The time averaged spectrum in the time interval $t_{GRB}$ has been fitted with a power law, obtaining a flux $(2.44\pm0.08) \times 10^{-7}$ erg cm$^{-2}$ s$^{-1}$ and a photon index $1.59\pm0.06$ in the 15-150 keV energy range. The fit was conducted using a time averaged DRM. The T$_{90}$ spectrum and the folded best fit model are shown in Figure \ref{fig:spe}.

\section{Conclusions}\label{Sec:conclusions}
We have introduced the updated \ba{} python package, which now includes the capability to analyze BAT TTE data that has been collected in multiple ways. Previously, the HEASoft tools allowed users to only analyze TTE data that was collected due to an on-board trigger. Now, the \ba{} software, allows users to process and analyze TTE data from on-board triggers, GUANO commanded TTE data dumps, and subthreshold triggers. The python package allows for more intuitive analysis of the data, and it conveniently integrates into the modern astrophysics python environment enhancing the use of BAT data in multi-wavelength and multi-disciplinary studies. We showcase the many features of the \ba{} software and verify its correctness by analyzing TTE data for GRB 050724, a short GRB with extended emission that triggered BAT. Furthermore, we show how the software can be used to analyze GUANO downlinked data, in the case of an FRB trigger for FRB 180916, a subthreshold TTE dataset obtained for an outburst of EXO 0748-676, and to localize and analyze the spectro-temporal properties of GRB 210706A, which occurred during a slew.

\subsection{Spectral Analysis Caveats}
One limitation of this software is that the spectral fitting convenience function included with \ba{} uses simple methods to fit spectra, which can sometimes give erroneous results. \added{These results can present as unconstrained spectral parameters with large errors associated with them. Many times, adjusting various pyXspec fitting parameters such as the number iterations that the code uses to converge on a solution or the permissible ranges on spectral model parameters can help with the fitting solution. We also would like to reinforce the need to utilize gaussian statistics when conducting spectral fitting with the mask-weighted BAT spectra, which is in contrast to other gamma-ray telescopes, and the need to ensure that the spectral models used are not overfitting the data. Additionally, it is prudent to ensure that the PHA and DRM files that are used in the fits are for as identical time periods as is possible to ensure that the behavior of the instrument is captured as accurately as possible in the DRM. The additional complication of constructing a time-averaged DRM for periods where Swift may be slewing is well worth the effort as the response of BAT can change over its large field of view. 

We also highlight that users can construct and use their own spectral fitting tools and spectral models within the greater python environment. This can take the form of models constructed with the astromodels package \citep{vianello_2021_5646925} or fitting algorithms that utilize scipy minimization functions \citep{scipy}, more rigorous monte carlo markov chain (eg. using the emcee package \citep{emcee}) or even nested sampling methods (eg using the MultiNest package \citep{multinest}), just to highlight a few. }

This work adds an important data analysis capability to the \ba{} software, unlocking the wealth of BAT TTE data for general use. The \ba{} software is now a comprehensive open-source pipeline for analyzing the most important scientific data products produced by BAT. The community can contribute to the open-source project by creating a fork on github and opening a pull request to incorporate their improvements into the official version of the Python package. Furthermore, the community can post any issues that they encounter to the github repository for expedited resolution of the problems.  This package may become an official part of HEASoft, in the future, making it an official Swift BAT data analysis pipeline.

\acknowledgements
The material is based upon work supported by NASA under award number 80GSFC21M0002. 

This material is based upon work supported by the National Aeronautics and Space Administration under Agreement No.80NSSC23K0552 issued through the Office of Science.  In accordance with Federal law, NMC is prohibited from discriminating on the basis of race, color, national origin, sex, age, or disability.

This research has made use of data and/or software provided by the High Energy Astrophysics Science Archive Research Center (HEASARC), which is a service of the Astrophysics Science Division at NASA/GSFC. We thank Bryan Irby and Abdu Zoghbi for help with HEASoftpy, Israel Martinez Castellanos for his development of Histpy, and Jamie Kennea for his assistance with swifttools.

\software{Astropy \citep{astropy:2013, astropy:2018,astropy:2022}, Astropy/reproject \citep{astropy_reproject},  Astroquery \citep{astroquery}, {BatAnalysis \citep{batanalysis}}
          NumPy \citep{numpy}, Matplotlib \citep{matplotlib}, Scipy \citep{scipy}, HEASoft \citep{HEASoft}, swiftbat\_python, Xspec \citep{xspec}, swifttools \citep{swifttools}, Histpy \citep{histpy_martinez_castellanos_2024_14262814}\\}

\bibliographystyle{aasjournal}
\bibliography{mybib}

\listofchanges

\end{document}